\documentclass[amssymb,prl,twocolumn,showpacs]{revtex4}
\usepackage{amsmath,amsthm,amsfonts}
\usepackage{amssymb}
\usepackage{graphicx}

\begin{document}
\title{
Shot Noise Spectrum of Open Dissipative Quantum Two-level Systems}
\author{Ram\'on Aguado $^1$ and Tobias Brandes $^2$}
\affiliation{1-Departamento de Teor\'{\i}a de la Materia Condensada,
Instituto de Ciencia de Materiales de Madrid,
CSIC, Cantoblanco 28049, Madrid, Spain}
\affiliation{2-Department of Physics, 
UMIST, P.O. Box 88, Manchester M60 1QD, United Kingdom}
\begin{abstract}
We study the current noise spectrum of {qubits} under transport conditions
in a dissipative bosonic environment. We combine
(non-)Markovian master equations with correlation functions in Laplace-space to
derive a noise formula for both weak and strong coupling to the bath.
The coherence-induced reduction of noise
is diminished by weak dissipation and/or a large {level separation (bias)}.
For weak dissipation, we demonstrate that the dephasing and relaxation rates of
the two-level system can be extracted from noise. In the
strong dissipation regime, the localisation-delocalisation transition
becomes visible in the low-frequency noise.
\end{abstract}
\date{\today{ }}
\maketitle
The way a quantum two-level system (qubit)
loses coherence due to the coupling with a noisy environment has
been the subject of intense research for many years
\cite{Legetal87,Weiss}.
This fundamental problem has received a great deal of
attention due to recent advances in solid state devices in which
quantum two-level systems (TLS) have been realized using different
degrees of freedom (charge, spin, flux) \cite{qubitexperiment}.
Interest in
current {\em noise} \cite{BB00}, in particular  in presence of dephasing and dissipation \cite{SU92},
has risen owing to the possibility of extracting
valuable information not available in conventional dc transport
experiments.

In this Letter, we demonstrate that current noise
in coupled quantum dots or Cooper pair boxes reveal
the complete dissipative, internal dynamics of qubits
coupled to external electron reservoirs.
We develop a formalism that allows us to make quantitative
predictions for the frequency ($\omega$) dependent
charge and current noise for {\em arbitrary} dissipative environments.
We find a
reduction of noise by coherent  oscillations, weakened by
increasing the bias or weak dissipation.
The latter suppresses shot noise at $\omega=0$ and large bias
due to spontaneous boson emission.
Importantly, the dephasing and relaxation rates of the TLS can be extracted from noise.
Our formulation
includes non-Markovian memory effects \cite{LD03} and
the strong coupling limit, where we observe
a re-establishing of the full shot noise
due to the formation of polarons as new quasi-particles.

In the following, we assume that the TLS is
defined in a double
quantum dot (DQD) device \cite{Fujetal98,Hayetal03}. We point out, however, that our method can also
be applied to charge qubits realized in a Cooper pair (CP) box \cite{MSS01,CGNS02,Choi03}, see below.
DQD´s in the regime of strong Coulomb
blockade can be tuned into a regime that is governed by a (pseudo)
spin--boson (SB) model (dissipative two--level system
\cite{Legetal87}), coupled to reservoirs\cite{BK99} ${\cal
H}={\cal H}_{SB}+{\cal H}_{res}+{\cal H}_{T}$. Here, ${\cal
H}_{SB}$ describes one additional `transport' electron which
tunnels between a left (L) and a right (R) dot with energy
difference $\varepsilon$ and inter--dot coupling $T_c$, and is
coupled to a dissipative bosonic bath (${\cal H}_B= \sum_{\bf
Q}\omega_{Q} a^{\dagger}_{\bf Q} a_{\bf Q}$),
\begin{equation}\label{modelhamiltonian}
{\cal H}_{SB}\!=\! \Big[\frac{\varepsilon}{2}
+\sum_{\bf Q} \frac{g_{Q}}{2} \left(a_{-\bf Q} + a^{\dagger}_{\bf
Q}\right)\Big]
\hat{\sigma}_z
\!+\! T_c
\hat{\sigma}_x
\!+\!{\cal H}_B.
\end{equation}
The effective Hilbert space of the closed system consists of two
states $|L \rangle=|N_L+1,N_R \rangle$ and $|R \rangle=|N_L,N_R+1
\rangle$, such that the system is defined by a `pseudospin' \cite{BK99}
$\hat{\sigma}_z \equiv |L \rangle \langle L|-|R \rangle \langle
R|\equiv \hat{n}_L-\hat{n}_R$ and $\hat{\sigma}_x\equiv |L \rangle \langle
R|+|R \rangle \langle L|\equiv \hat{p}+\hat{p}^{\dagger}$. The effects
of the
bath can be encapsulated in the spectral density
$J({\omega})\equiv\sum_{\bf Q} |g_{Q}|^2\delta(\omega-\omega_Q)$,
where $\omega_Q$ are the frequencies of the bosons and the $g_Q$
denote interaction constants.
{ When showing
results we will be using
$J(\omega)={2\alpha}{\omega}
\left[1-{\omega_d}/{\omega}\sin\left({\omega}/{\omega_d}\right)\right]
e^{-\omega/\omega_c}$ for piezoacoustic phonons in lateral DQD´s
with $\omega_d$ depending on the
geometry \cite{BK99}, or a generic Ohmic bath $(\omega_d\to 0)$}:
$J(\omega)={2\alpha}\omega e^{-\omega/\omega_c}$. The
dimensionless parameter $\alpha$ reflects the strength of
dissipation and $\omega_c$ is a high energy cutoff \cite{Weiss}.
The coupling to
reservoirs ${\cal H}_{res}=\sum_{k_\alpha}\epsilon_{k_\alpha}
c_{k_\alpha}^{\dagger}c_{k_\alpha}$
is described by ${\cal H}_T=\sum_{k_\alpha{}}
(V_k^\alpha{} c_{k_\alpha{}}^{\dagger}s_\alpha{}+H.c.)$, where
$\hat{s}_\alpha=|0 \rangle
\langle \alpha|$ ($\alpha$=L,R) and
the extra
state $|0 \rangle=|N_L,N_R \rangle$ describes an `empty' DQD, such that
$1=\hat{n}_0+\hat{n}_L+\hat{n}_R$.

The full model described by ${\cal H}$
allows to study
non--equilibrium properties, such as the inelastic stationary
current or current noise, through an open dissipative TLS.
We describe its dynamics by a reduced, with respect to reservoirs,
statistical operator $\rho(t)$.
Introducing the vectors ${\bf A}\equiv
(\hat{n}_L,\hat{n}_R,\hat{p},\hat{p}^{\dagger})^T$, ${\bf
\Gamma}=(\Gamma_L,0,0,0)^T\equiv\Gamma_L{\bf e}_1$
and a matrix memory kernel ${M}$,
the equations of motion (EOM) of the expectation values \cite{BV02}
(with $\langle \hat{O} \rangle\equiv \sum_{i=0,L,R}{\rm Tr_{bath}} \langle i |\hat{O}\rho(t)|i\rangle$)
read in matrix form
\begin{equation}\label{expectation}
  \langle {\bf A}(t)\rangle   \!=\! \langle {\bf A}(0)\rangle
 + \int_{0}^{t}dt' \left\{{M}(t-t')\langle {\bf A}(t')\rangle  + {\bf \Gamma} \right\}.
\end{equation}
Eq.~(\ref{expectation}) can be 
solved in Laplace space as
$\langle\hat{\bf A}(z)\rangle= [z-z\hat{M}(z)]^{-1}(\langle{\bf
A}(0)\rangle+{\bf \Gamma}/z)$ and serves as a starting point for the
analysis of stationary ($1/z$ coefficient in Laurent series for $z\rightarrow
0$) and non-stationary quantities. The memory kernel has a
block structure
\begin{equation}\label{block}
  z\hat{M}(z)=\left[
  \begin{matrix}
    -\hat{G} & \hat{T}_c\\ \hat{D}_{z}& \hat{\Sigma}_{z}
  \end{matrix}\right], \quad
 \hat{G}\equiv
\left(
  \begin{matrix}
\Gamma_L & \Gamma_L\\0&\Gamma_R
\end{matrix}\right),
\end{equation}
where $\hat{T}_c \equiv -iT_c(1-\sigma_x)$,
and the coupling to the reservoirs within Born and Markov (BM) approximation with respect to ${\cal H}_{T}$
\cite{BK99,SN96} is given by
$\Gamma_{\alpha}=2\pi\sum_{k_{\alpha}}|V_k^{\alpha}|^2\delta(\epsilon-\epsilon_{k_{\alpha}})$
(we assume Fermi distributions for the reservoirs $f_L=1$ and
$f_R=0$; large voltage regime).
The blocks $\hat{D}_{z}$ and $\hat{\Sigma}_{z}$
are determined by the EOM for the coherences (off--diagonal elements)
$\langle\hat{p}\rangle=\langle\hat{p}^{\dagger }\rangle^*$
and contain the complete information on dephasing of the system.
In general, no exact solution is available
but we will present approximate results now:
for weak coupling
to the bosons,  one can use perturbation theory (PER)
in $\alpha$
in the correct basis of the hybridised states of the TLS.
In BM approximation, the resulting expressions are:
\begin{equation}\label{PERb}
  \hat{D}^{\rm PER}=\hat{T}_c+
\left(
 \begin{matrix}
\gamma_+ &  - \gamma_-\\ \gamma_+ &  -\gamma_-
\end{matrix}\right),\quad
\hat{\Sigma}^{\rm PER}=
\left(
 \begin{matrix}
E&0\\0&E^*
\end{matrix}\right),
\end{equation}
where $E=i\varepsilon-\gamma_p-\frac{\Gamma_R}{2}$,
$\gamma_p\equiv 2\pi \frac{T_c^2}{\Delta^2} J(\Delta) \coth
\left(\beta \Delta /2\right)$ and
$\gamma_{\pm}\equiv-\frac{\varepsilon T_c}{\Delta^2} \frac{\pi}{2}
J(\Delta)  \coth \left(\beta \Delta /2\right)\mp
 \frac{T_c}{\Delta} \frac{\pi}{2}J(\Delta)$
completely determine dephasing and relaxation in the system.
Here, $\Delta\equiv\sqrt{\varepsilon^2+4T_c^2}$ is the
hybridization splitting and $\beta=1/k_BT$.

On the other hand,
for strong electron-boson coupling, one has to start from a
polaron--transformed frame (strong coupling, POL), leading to an
integral equation \cite{BK99} which involves the boson correlation
function $C(t)\equiv\exp({-\int_0^{\infty}d\omega
\frac{J(\omega)}{\omega^2} \left[ \left(1- \cos \omega t\right)
\coth \left(\frac{\beta \omega}{2}\right) + i \sin \omega t
\right]})$. Introducing
$C^{[*]}_\varepsilon(z)\equiv\int_{0}^{\infty}dte^{-zt}e^{[-]i\varepsilon
t} C^{[*]}(t)$, the resulting matrices in $z$-space are
\begin{equation}\label{POLb}
   \hat{D}_{z}^{\rm POL}=iT_c
\left(\begin{matrix}
-1 & {\hat{C}^*_{-\varepsilon}}/{\hat{C}_\varepsilon}\\
1 &  -{\hat{C}_{-\varepsilon}}/{\hat{C}^*_\varepsilon}
\end{matrix}\right),
\hat{\Sigma}_{z}^{\rm POL}=
\left(
 \begin{matrix}
\tilde{E}&0\\0&\tilde{E}^*
\end{matrix}\right),
\end{equation}
with $\tilde{E}^{[*]}\equiv z-1/C_\varepsilon^{[*]}(z)-\Gamma_R/2$.
In contrast to the PER solution, where
$M(\tau)=M=z\hat{M}(z)$ is time-independent,
$M^{\rm POL}(\tau)$ is time-dependent and
$z\hat{M}(z)$ depends on $z$ in the POL approach \cite{footnote2}.
We note
that $\mbox{Re}[C_\varepsilon (z)]|_{z=\pm i\omega}=\pi
P(\varepsilon\mp\omega)$
where $P(\varepsilon)$ is the probability for inelastic tunneling
with energy transfer $\varepsilon$ \cite{Weiss}.

As mentioned above, our model describes
a CP box as well, the transport through the DQD
being analogous to the Josephson Quasiparticle Cycle (JQP) of
the superconducting single electron transistor (SSET) with $E_C\gg E_J$,
such that only two charge states,
$|2\rangle$ (one excess CP
in the SSET) and $|0\rangle$ (no extra CP), are allowed. Two
consecutive quasiparticle events (with rates $\Gamma_{2}$ and
$\Gamma_{1}$) couple
$|2\rangle$ and $|0\rangle$ with another state $|1\rangle$ through
the cycle $|2\rangle\rightarrow |1\rangle\rightarrow
|0\rangle\Leftrightarrow |2\rangle$.
Interdot tunneling is analogous to coherent tunneling of a
CP through one of the junctions, and tunneling to and
from the DQD
is analogous to the two
quasiparticle events through the probe junction in the SSET \cite{Choi03}.

Current noise, which is described by the power spectral density $\mathcal
{S}_{I}(\omega)\equiv 2\int_{-\infty}^{\infty} d\tau
e^{i\omega\tau} \mathcal {S}_{I}(\tau)=\int_{-\infty}^{\infty}
d\tau e^{i\omega\tau}\langle
\{\Delta\hat{I}(\tau),\Delta\hat{I}(0)\} \rangle$, with $ \Delta\hat{I}(t)\equiv \hat{I}(t)-\langle
\hat{I}(t)\rangle$, is
a sensitive tool to study correlations between carriers \cite{BB00}.
The Fano factor ($\gamma\equiv\frac{\mathcal{S}_I(0)}{2qI}$)
quantifies deviations from the Poissonian noise,
$\mathcal{S}_I(0)=2qI$, which characterizes uncorrelated carriers
with charge $q$. Importantly, $S_I(\omega)$ has to be calculated
from the autocorrelations of the {\it total} current $I(t)$, i.e.
particle plus displacement current \cite{BB00}. Using current
conservation together with the Ramo-Shockley theorem, $I(t)=a
I_L(t) + b I_R(t)$ ($a$ and $b$, with $a+b=1$, depend on each
junction capacitance \cite{BB00}), one can express $\mathcal
{S}_{I}(\omega)$ in terms of the spectra of particle currents and
the charge noise spectrum $S_{Q}(\omega)$ \cite{Mozetal02},
\begin{equation}
\label{fullnoise}
 S_I(\omega)=a S_{I_L}(\omega) + b
 S_{I_R}(\omega)-a b\omega^2S_{Q}(\omega).
\end{equation}
Note that in symmetric configurations, $a\approx b$, the charge
noise reduces the contribution from particle currents to the noise
spectrum. For $a\approx 1$ or $b\approx 1$ the main contribution
to noise comes from particle currents. At zero frequency
$S_I(0)=S_{I_L}(0)=S_{I_R}(0)$.
$S_{Q}(\omega)$ is defined as
\begin{eqnarray}
\label{charge}
  S_Q(\omega) &\equiv& \lim_{t\to \infty}
\int_{-\infty}^{\infty} d\tau e^{i\omega\tau}\langle
\{\hat{Q}(t),
\hat{Q}(t+\tau)\}\rangle \nonumber\\
&=& 2{\mbox Re} \left\{ \hat{f}(z=i\omega) +
\hat{f}(z=-i\omega)\right\},
\end{eqnarray}
where $\hat{Q}=\hat{n}_L+\hat{n}_R$ and $\hat{f}(z)$ is the
Laplace transform of
\begin{eqnarray}
\label{fequation}
  f(\tau) &=& \sum_{i,j=L,R}\langle \hat{n}_i(t)\hat{n}_j(t+\tau)\rangle
\end{eqnarray}
and can be evaluated with the help of the charge correlation
functions ${\bf C}_\alpha(\tau)\equiv \langle \hat{n}_{\alpha}(t) {\bf
A}(t+\tau) \rangle$, as $f(\tau)=({\bf e}_1+{\bf
e}_2)[{\bf C}_L(\tau)+{\bf C}_R(\tau)]$. The EOM for ${\bf
C}_\alpha(\tau)$ can be obtained from the quantum regression theorem \cite{Carmichael}
whose solution is again expressed with the help of the resolvent $[z-z\hat{M}(z)]^{-1}$, cf. Eq.
(\ref{block}).
\par
\begin{figure}[t]
\includegraphics[width=0.8\columnwidth]{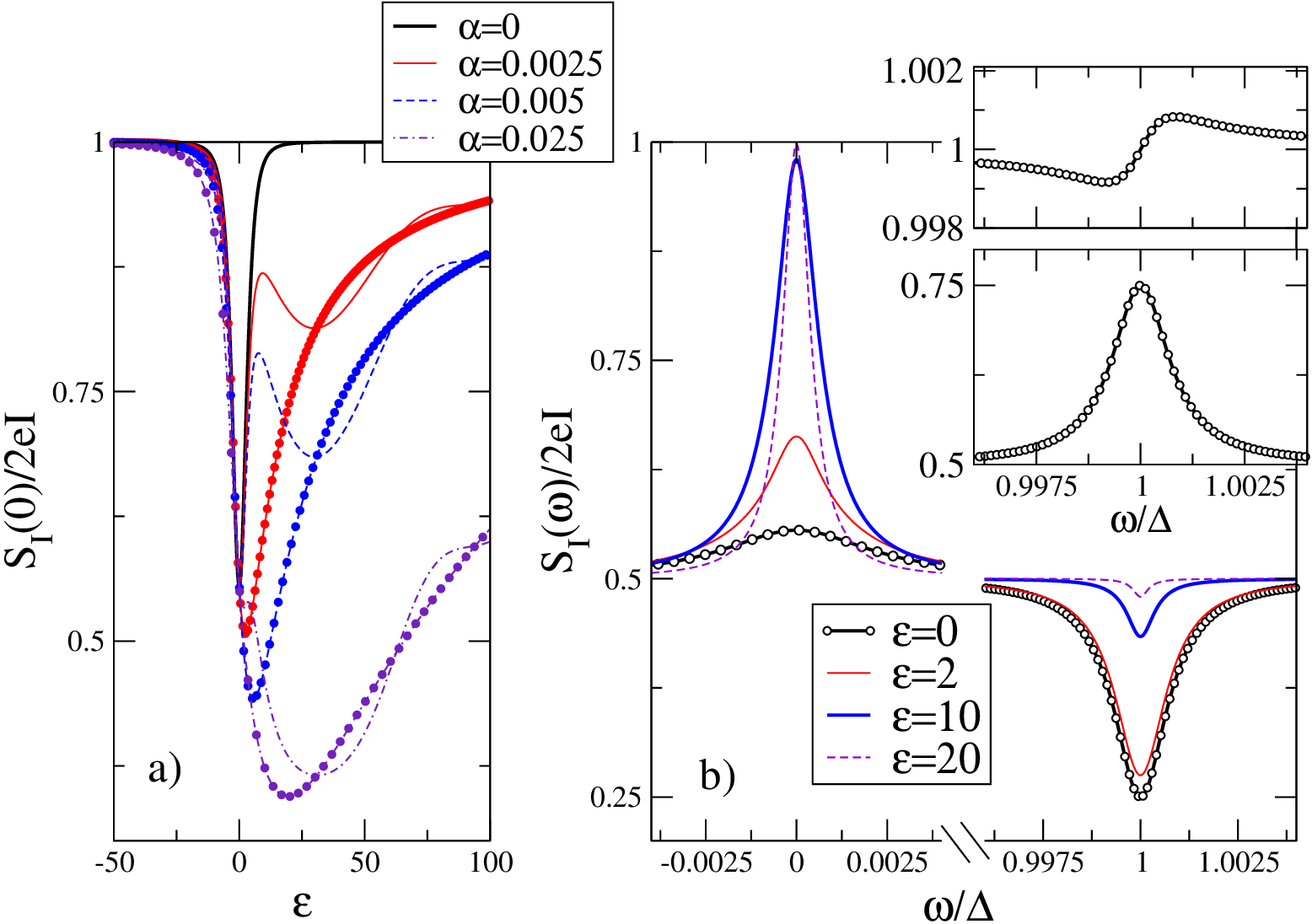}
\caption[]{a) Fano factor vs. bias $\varepsilon$ for
different dissipative couplings $\alpha$.
Parameters $T_c=3$, $\Gamma=0.15$, $\omega_c=500$,
$\omega_d=10$, $T=2$ (in $\mu eV$) correspond to
typical experimental values~\cite{Fujetal98} in double quantum dots.
Lines: acoustic phonons, circles: generic ohmic environment $\omega_d=0$ (see text).
b) Frequency dependent current noise ($\alpha=0$, $T=0$, $\Gamma=0.01$).
Inset: (Top) Contribution to noise from particle currents $S_{I_R}(\omega)/2eI$. (Bottom) Charge noise contribution $\omega^2S_{Q}(\omega)/8eI$. $a=b=1/2$.} 
\end{figure}
To calculate the contribution of particle currents to noise, we need to relate the
reduced dynamics of the qubit described by Eqs. (2-3) to reservoir
operators.
For $S_{I_R}(\omega)$, 
we introduce
the number $n$ of electrons that have
tunneled through the right barrier \cite{MSS01,EG02,RK03})
which defines generalized expectation values as
$O^{(n)}\equiv\sum_{i=0,L,R}{\rm Tr_{bath}} \langle n,i |\hat{O}\rho(t)|n,i\rangle$
(such that $\langle \hat{O} \rangle=\sum_n O^{(n)}$)
and write
\begin{eqnarray}\label{generalized}
  \dot{n}_0^{(n)}&=&-\Gamma_L {n}_0^{(n)} + \Gamma_R {n}_R^{(n-1)}\nonumber\\
  \dot{n}_{L/R}^{(n)}&=& \pm\Gamma_{L/R} {n}_0^{(n)} \pm iT_c \left( p^{(n)}-
[p^{(n)}]^{\dagger}\right)
\end{eqnarray}
and correspondingly for
$p^{(n)}$ and $[p^{(n)}]^{\dagger}$
and the left barrier.
Eqs.(\ref{generalized}) allow one to calculate the particle current
and the noise spectrum from
$P_n(t)=n_0^{(n)}(t)+n_L^{(n)}(t)+n_R^{(n)}(t)$ which gives the
total probability of finding $n$ electrons in the collector by time
$t$. In particular, $I_R(t)=e\sum_n n\dot{P}_n(t)$
and $S_{I_R}$ can be calculated from \cite{Mac48}
\begin{eqnarray}
  S_{I_R}(\omega)=2\omega e^2\int_0^{\infty} dt \sin(\omega t) \frac{d}{dt}
\left[ \langle n^2(t) \rangle - (t\langle I\rangle)^2 \right],
\end{eqnarray}
where $\frac{d}{dt}\langle n^2(t)\rangle=\sum_n
n^2\dot{P}_n(t)=\Gamma_R\sum_{n=0}^{\infty} n {n}_R^{(n)} (t) +
\Gamma_R \sum_{n=0}^{\infty} {n}_R^{(n)} (t)$.
Solving Eqs.(\ref{generalized}) with the initial condition
$n_R^{(n)} (0)=\delta_{n,0}n_R(0)$, where $n_R(0)$ is the
stationary solution of Eqs. (2) \cite{RK03}, we get
\begin{eqnarray}\label{SIR}
S_{I_R}(\omega) &=& 2eI\left\{1+\Gamma_R\left[\hat{n}_R(-i\omega)
+ \hat{n}_R(i\omega)\right]\right\},
\end{eqnarray}
with
$z\hat{n}_R(z)=\Gamma_Lg_+(z)/N(z)$,  where
$N(z)\equiv[z+\Gamma_R+g_-(z)](z+\Gamma_L)$
$+(z+\Gamma_R+\Gamma_L)g_+(z)$ and
\begin{eqnarray}\label{gpm}
 g_{+[-]}(z)=\pm iT_c({\bf e}_1-{\bf e}_2)
 \left[z-\hat{\Sigma}_z\right]^{-1}\hat{D}_z {\bf e}_{1[2]}.
\end{eqnarray}
Eqs. (\ref{SIR},\ref{gpm}) demonstrate
the dependence of the current noise on the dephasing via the two-by-two blocks $\hat{D}_z$ and $\hat{\Sigma}_z$,
cf. Eq. (\ref{block},\ref{PERb},\ref{POLb}).
Explicitly,
\begin{eqnarray}\label{gpm1}
g^{\rm PER}_{\pm}(z)&\equiv&
  2T_c\frac{T_c(\gamma_p+ \Gamma_R/2+z)-\varepsilon\gamma_{\pm}}{(\gamma_p+ \Gamma_R/2+z)^2+\varepsilon
  ^2}\nonumber\\
g^{\rm POL}_{+[-]}(z) &\equiv& T_c^2 \left[
\frac{C^{[*]}_{[-]\varepsilon}(z)}{1+\frac{\Gamma_R}{2}C_\varepsilon(z)}+
(C \leftrightarrow C^*)
\right].
\end{eqnarray}
A similar derivation yields $S_{I_L}(\omega)=S_{I_R}(\omega)$. 
The explicit expressions
Eq.(\ref{SIR},\ref{gpm},\ref{gpm1}), together with the inverse of a 4 by 4 matrix for the charge noise
Eqs.~(\ref{charge}), yield our key quantity $S_I(\omega)$,  Eq.~(\ref{fullnoise}).

{\it Zero frequency (Shot Noise)}.--
In the zero frequency limit
$z\rightarrow 0$,
one obtains
\begin{eqnarray}
\label{shot}
  S_I(0) = 2eI \left(1+2\Gamma_R \frac{d}{dz}\left[z
  \hat{n}_R(z)\right]_{z=0}\right).
\end{eqnarray}
Eqs.~(\ref{shot})
allows to
investigate the shot noise of open dissipative TLS's
for {\it arbitrary environments}.  In contrast to
non-interacting mesoscopic conductors, the noise cannot be
written in the Khlus-Lesovik form $S_I(0)=
2 e^2 \int \frac{dE}{2\pi} t(E) [1-t(E)]$ with an effective
transmission coefficient $t(E)$.
Without bath,
we recover the results of Ref.~[\onlinecite{EG02}] (shot noise
of DQD's) and Ref.~[\onlinecite{Choi03}] (shot noise
of the CP box). For $\alpha=0$ and
$\Gamma\equiv\Gamma_L=\Gamma_R$ (Fig. 1a, solid line), the
smallest Fano factor is reached for $\varepsilon =0$ where quantum
coherence strongly suppresses noise. The maximum suppression
($\gamma=1/5$) is reached for $\Gamma =2\sqrt{2}T_c$. For large
$\varepsilon>0$ ($\varepsilon <0$) the charge becomes localized in
the right (left) level, $S_I(0)$ is dominated by only one Poisson
process, namely the noise of the right (left) barrier, and
$\gamma\rightarrow 1$.
\begin{figure}[t]
\begin{center}
\includegraphics[width=0.7\columnwidth]{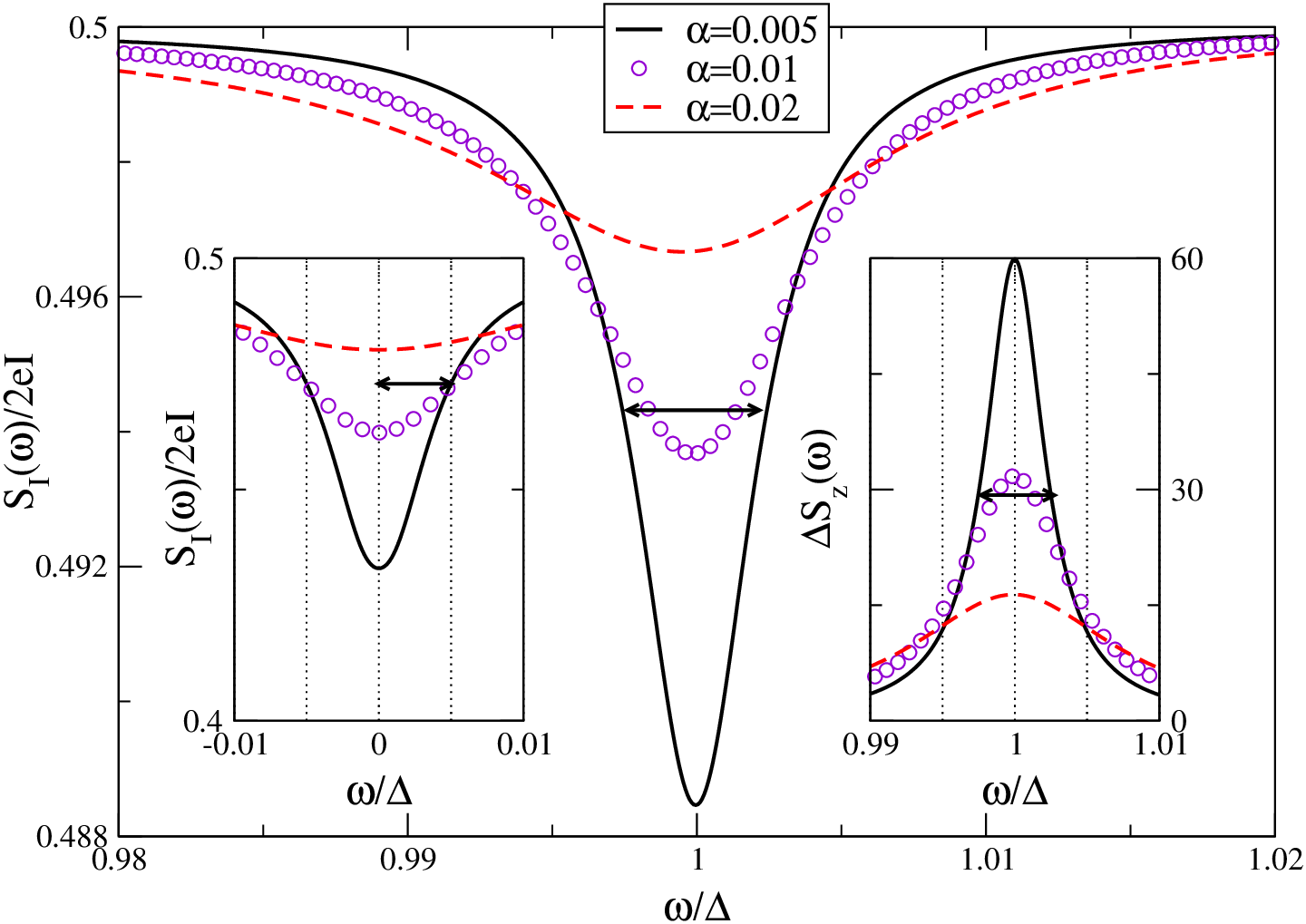}
\end{center}
\caption[]{Effect of ohmic dissipation on current noise near resonance
($\varepsilon=10$, $\Gamma=0.01$,
and  $\alpha=0.005, 0.01, 0.02$
corresponding to $\gamma_p\approx 4.74\Gamma, 9.47\Gamma, 18.95\Gamma$).
Inset: (Right)
pseudospin correlation function $S_z(\omega)$. 
Arrows indicate 
relaxation rate
$(\Gamma+\gamma_p)/\Delta\approx 0.005$ for $\alpha=0.005$.
(Left)
low frequencies region near shot noise limit $\omega=0$. 
}
\end{figure}
For $\alpha\neq 0$, this mechanism is strongly
affected by the possibility of exchanging energy quanta with the
bath.
The effect of a bosonic bath on noise in a mesoscopic scatterer was first discussed in
\cite{US93}. For the TLS discussed here, spontaneous
emission (for $\varepsilon >0$) occurs even at very low temperatures \cite{Fujetal98,BK99}
and the noise is
reduced \cite{SU92} well below the Poisson limit (Fig. 1a).
The maximum suppression is now reached when the elastic and inelastic rates coincide, i.e.,
$\gamma_p=\Gamma_R$, as we have checked numerically.
For large
couplings, spontaneous emission leads to a very asymmetric Fano factor that goes
from $\gamma\approx 1$ to $\gamma\approx 0.5$ as $\varepsilon$
changes sign (not shown here).

{\it Finite frequencies}.--
For finite $\omega$, 
the numerical results for
$\alpha=0$ are shown in Fig. 1b where we plot $S_I(\omega)$ for different
values of $\varepsilon$. The background noise is half the Poisson
value as one expects for a symmetric structure. $\gamma$ deviates
from this value around $\omega=0$
where the noise has a peak
and $\omega=\Delta$ where the
noise is suppressed. The dip in the Fano factor directly reflects the
resonance  of the subtracted charge noise $S_{Q}(\omega)$ around $\Delta$ (inset Fig. 1b), cf. Eq.
(\ref{fullnoise}).
An increase of $\varepsilon$ localizes the qubit and,
thus, the zero-frequency noise reaches $\gamma\rightarrow 1$.
Moreover, the dip in the high frequency noise at $\omega=\Delta$
(Fig. 1b) is progressively destroyed (reduction of quantum
coherence) as $\varepsilon$ increases which is consistent with the
previous argument.
\begin{figure}[t]
\begin{center}
\includegraphics[width=0.7\columnwidth]{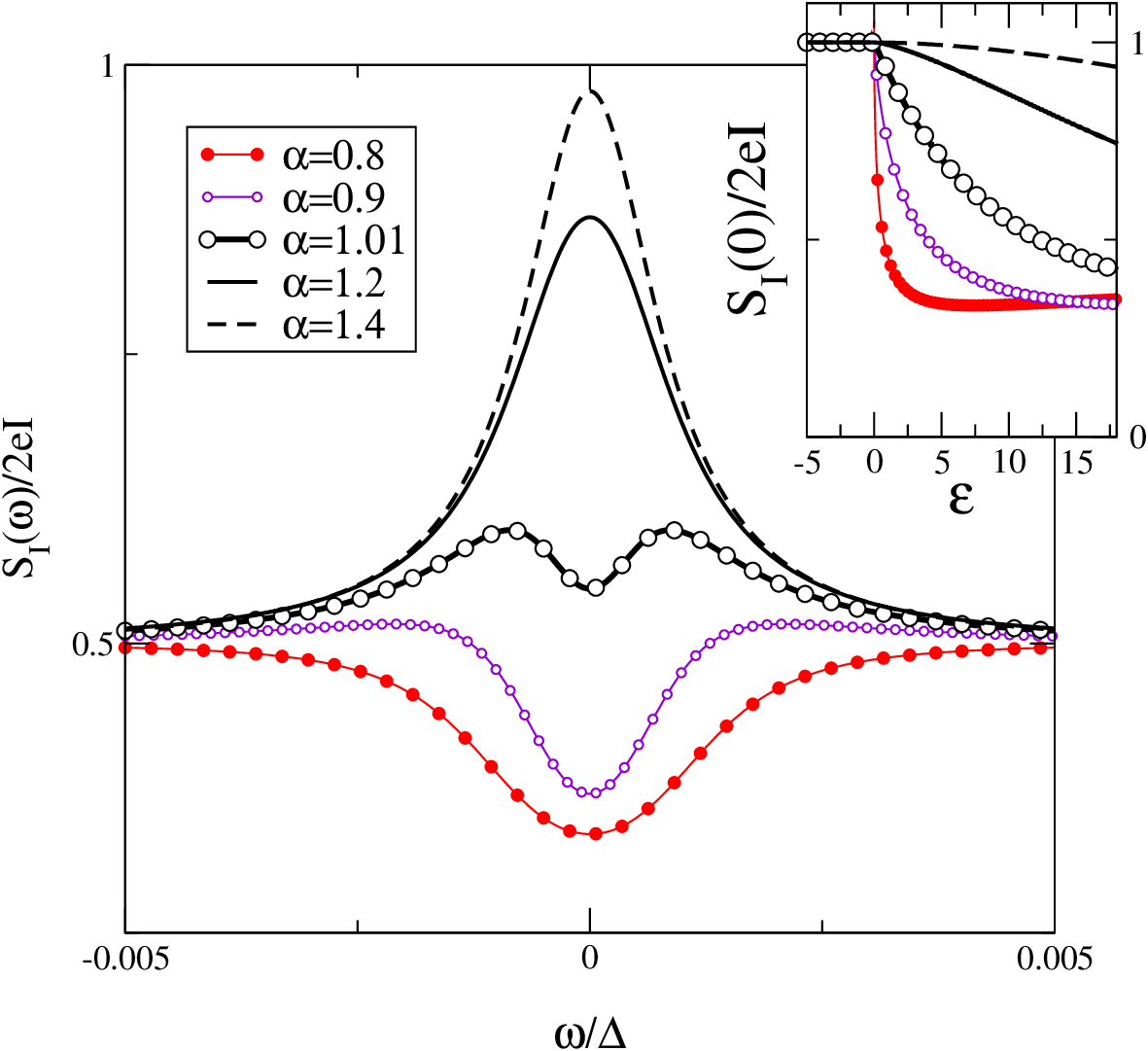}
\end{center}
\caption[]{Low frequency current noise in qubit with strong ohmic dissipation
($T=0$, $\varepsilon=10$, $T_c=3$, $\Gamma_L=\Gamma_R=0.01$).
Inset:  Shot noise for $\omega=0$.
}
\end{figure}
A similar reduction of the dip at $\omega=\Delta$ occurs at
fixed $\varepsilon$ and $\Gamma$ with increasing dissipation
(Fig 2) in the weak coupling (PER) regime. This behavior demonstrates
that $S_I(\omega)$ reveals the complete internal dissipative dynamics
of the TLS.
The above argument can be further substantiated by
plotting also the symmetrized pseudospin
correlation function $S_z(\omega)=1/2\int_{-\infty}^{\infty} d\omega
e^{i\omega\tau}\langle\{\hat{\sigma}_z(\tau),\hat{\sigma}_z\}\rangle$ (Fig. 2, right inset)
which is commonly used to investigate the dynamics of the SB problem \cite{Weiss}.
Both functions reflect in the same fashion how the coherent dynamics of the system progressively
gets damped by the bosonic bath.

In particular, the {\em dephasing rate} can be extracted from the half-width of
$S_I(\omega)$ around
$\omega=\Delta$. For an Ohmic environment, $\gamma_d^b=\gamma_p/2+2\pi\alpha(\frac{\varepsilon}{\Delta})^2 k_BT$, such that
the total dephasing rate is
$\gamma_d(T=0)=\gamma_d^b+\Gamma/2=(\gamma_p+\Gamma)/2$ (Fig. 2, arrows denote full-width, i.e.
$2\gamma_d\approx \gamma_p$ as $\alpha$ increases).
Close to $\omega=0$, the peak in $S_I(\omega)$ for $\alpha=0$
changes into a dip around
$\omega=0$ reflecting incoherent relaxation dynamics for $\alpha\neq 0$.
The half-width is now given by the {\em relaxation rate} such that
the full-width of $S_I(\omega)$ around $\omega=0$ is
{\it twice} that of the high frequency noise (Fig 2, left inset).

The results for the strong coupling (POL) regime are presented in Fig 3.
Near $\omega=0$,
POL and PER yield nearly identical results for the noise $S_I(\omega)$
at very small $\alpha$ (not shown here).
The cross-over to Poissonian noise near $\omega=0$ with increasing $\alpha$ indicates
the formation of localized polarons. The delocalisation-localisation transition \cite{Legetal87,Weiss}
of the spin-boson model at $\alpha=1$  is reflected in a
change of the analyticity of $C_\varepsilon$ and the shot noise near zero bias
(Fig 3, inset).
Similar physics has been found recently in
the suppression of the persistent current $I(|\varepsilon|)\propto
\mbox{\rm Im}C_{-|\varepsilon|}$ through
a strongly dissipative quantum ring containing  a quantum dot with bias $\varepsilon$
\cite{CB01}. Although POL becomes less reliable for $\alpha<1$ and smaller bias,
the non-symmetry in $\varepsilon$ of the shot noise and the inelastic current $\propto
\mbox{\rm Re} C_\varepsilon$ reflects the
`open' topology of our TLS in the non-linear transport regime.

To conclude, our results  demonstrate that frequency-dependent current noise
provides detailed information about the internal, dissipative dynamics
of open quantum two-level system such as
double quantum dots or Cooper pair boxes. The
weak coupling regime should be close to current experiments \cite{Leonoise}
in these systems, where
we expect our predictions to be tested  in the near future.

We thank Markus B\"uttiker, Leo P. Kouwenhoven and Till Vorrath for useful discussions.
This work was supported by
EPSRC GR/R44690, DFG BR 1528 and by the
MCYT of Spain through the "Ram\'on y Cajal" program and grant MAT2002-02465 (R. A.).



\begin{thebibliography}{20}
\bibitem{footnote2}
For the SB problem,
POL is equivalent to the double-path integral non-interacting
blip approximation \cite{Legetal87,Weiss} and works well for $\varepsilon=0$
and $\varepsilon\gg T_c$.
\end{thebibliography}


\end{document}